\documentclass[preprint,nofootinbib]{revtex4}%
    \usepackage{amssymb}
    \usepackage{amsfonts}
    \usepackage{amsmath}
    \usepackage[utf8]{inputenc}
    \usepackage[OT1,T2A]{fontenc}
    \usepackage{amsmath}
    \usepackage{amsfonts}
     \usepackage{multirow}
    \usepackage{graphicx}
    \usepackage{bm}
    \usepackage{tikz}
    \usetikzlibrary{cd}
    \usepackage{musicography}
    \usepackage{tcolorbox}
    \usepackage{wrapfig}
    \usepackage{tensor}
    \usepackage{fancyhdr}
    \usepackage{hyperref}
    \usepackage{halloweenmath}
    \usepackage{comment}
    \usepackage{xcolor}
    \usepackage{cancel}
    \usepackage{wrapfig}
    \providecommand{\U}[1]{\protect\rule{.1in}{.1in}}
    \hypersetup{colorlinks,linkcolor={blue},citecolor={blue},urlcolor={black}} 
    \providecommand{\U}[1]{\protect\rule{.1in}{.1in}}
    \definecolor{blue}{rgb}{0,0,1}
    
    \definecolor{red}{rgb}{1,0,0}
    
\begin{document}
    \title{The Carrollian limit of ModMax electrodynamics}
    \author{$^1$Francisco Correa, $^2$Ankai Hern\'{a}ndez, $^1$Julio Oliva}
    \let\OLDthebibliography\thebibliography
\renewcommand\thebibliography[1]{
  \OLDthebibliography{#1}
  \setlength{\parskip}{0pt}
  \setlength{\itemsep}{-0.3ex}
}

    \affiliation{$^1$Departamento de Física
    Universidad de Santiago de Chile, Av. Victor Jara 3493, Santiago, Chile.}

    \affiliation{$^2$Departamento de F\'{\i}sica, Universidad de Concepci\'{o}n, Casilla, 160-C,
    Concepci\'{o}n, Chile.}

    \begin{abstract}
    We consider the Carrollian limit of ModMax electrodynamics, namely the limit of vanishing
    speed of light, for the most general, four-dimensional, duality and conformal
    invariant electromagnetism. The theory is parameterized by a unique real
    constant $\gamma$, which remains playing a non-trivial role in the magnetic
    Carrollian case, while it can be removed in the electric Carrollian contraction, and we therefore focus in the former. Applying the technique of Lie point symmetries, we obtain that the 
    magnetic limit is invariant under the Carrollian group, as well as under the local
    translation in Carrollian time $x^{0}\rightarrow x^{0\prime}=x^{0}+f\left(
    x^{i}\right)  $ and $x^{i}\rightarrow x^{i\prime}=x^{i}$, with $f$ being an
    arbitrary function. A diagonal part of the symmetries span the Conformal Carroll algebra of level $2$, $\mathfrak{ccarr}_2$ in four dimensions. Two additional internal symmetries remain in the Carrollian limit of ModMax standing for the conformal invariance of the theory, as well as the invariance under duality transformations.
    \end{abstract}
    \maketitle

    \section{Introduction}
    
    The analysis of possible kinematic groups by L\'evy-Leblond and
    Bacry \cite{bll} leads to the Carroll algebra, which was originally discovered in \cite{ll} and \cite{sen}. This algebra can be constructed from
    the Poincare algebra by taking the contraction $c\rightarrow0$, after a suitable rescaling of the
    generators. The algebra remained a curiosity
    for many years, due to the non-standard properties that a physical system realizing this algebra as a symmetry algebra must have. For example, when
    taking the $c\rightarrow0$ limit in the Minkowski metric, the light cones
    collapse making impossible the motion for a non-tachyonic particle. 
    %{\bf (for some of the sporadic, but relevant works, where this algebra was considered see [asdf] and references therein)}.
    In the last decade there has been a revival of interest in the Carroll algebra, since it was realized that one of its conformal extensions matches precisely with the BMS algebra of Bondi-Metzner-Sachs.  Such algebra controls the asymptotic structure of General Relativity in asymptotically flat spaces as one approaches null-infinity \cite{bms} and it is therefore relevant to extensions of holography \cite{malda, gkp, witten} to the vanishing cosmological constant case (see e.g. \cite{Donnay:2023mrd} and references therein). The properties of the Carroll algebras and their emergence as symmetries of physical systems have been studied in different scenarios, including for example the Maxwell theory, their relation to gravitational waves as well as in higher spin theories in $2+1$ dimensions  \cite{d1,d2,conformal,bgl, grav,Concha:2021jnn,Concha:2022muu}.
    
    In particular, the $c\rightarrow0$ limit of nonlinear electrodynamics has been recently addressed, exploring the consequences of Carrollian symmetries at the level of self-interacting extensions of electromagnetism. In the case of Born-Infeld theory, it has been argued in \cite{borncar} that the presence of dimensional couplings, which remain in the Carrollian limit, does not allow for the realization of the conformal Carrollian symmetry. The recent discovery of a nonlinear extension of electromagnetism invariant under conformal and duality transformations, called ModMax theory \cite{ModMax}, raised the natural question of the role of Carrollian symmetries in this construction. Thus, electric Carrollian field theories that are ModMax-like were recently built in \cite{sdca}. The construction makes use of a Carrollian Hodge
    dualization, which is subtle since the Carrollian structure contains a
    degenerate metric. This new operation permits to implement a generalization of
    electromagnetic duality at the level of the non-linear, Carrollian
    electrodynamics. On the other hand, the Galilean limit of ModMax theory has been explored in \cite{gaMM}. Our approach in the present work differs from the latter in that we consider the
    $c\rightarrow0$ limit directly at the level of the field equations of the
    original ModMax theory, and then we construct the symmetry generators using the tool of Lie point symmetries. In our approach, the infinite
    dimensional point symmetry $x^{0}\rightarrow x^{0\prime}=x^{0}+f\left(
    x^{i}\right)$ emerges naturally, as well as the conformal and duality
    invariance. Our paper is organized as follows. In Section \ref{two} the ModMax theory is briefly defined and both electric and magnetic Carrollian limits are presented. We also provide an exact solution of the non-trivial magnetic Carrollian ModMax theory, which explicitly depends on the ModMax coupling $\gamma$. Section \ref{three} is devoted to the analysis of the symmetries present in Carrollian ModMax theory. In Section \ref{four} we present some final remarks and future prospects.

    \section{ModMax electrodynamics and its Carrollian limits}\label{two}
    
    Modified Maxwell theory (ModMax theory) was discovered in \cite{ModMax}, during a thorough exploration of non-linear
    electromagnetism with Lagrangians $\mathcal{L}\left(S,P\right)$ depending on both
    relativistic %(duality)
    invariants
    \begin{align}
    S&=-\frac{1}{4}F^{\mu\nu}F_{\mu\nu}=\frac{1}{2}\left(\frac{E^{2}}{c^{2}}-B^{2}\right)\ , \\
    P&=-\frac{1}{4}\bar{F}^{\mu\nu}F_{\mu\nu}=\frac{1}{c}\bm{B}\cdot\bm{E}\ . 
    \end{align}
    In terms of these invariants, the most general Lagrangian invariant under conformal and duality transformations can be characterized by a real parameter $\gamma$ and reads
    \begin{align}
    	\mathcal{L}_{\gamma}:&=\cosh\gamma\:S+\sinh\gamma\sqrt{S^{2}+P^{2}}\\
    	&=\frac{1}{2}\cosh\gamma\left(\frac{E^{2}}{c^{2}}-B^{2}\right)+\sinh\gamma\sqrt{\frac{1}{4}\left(\frac{E^{2}}{c^{2}}-B^{2}\right)^{2}+\frac{1}{c^{2}}\left(\bm{E}\cdot\bm{B}\right)^{2}}\ .
    \end{align}
    By construction, the Lagrangian preserves the main properties of Maxwell theory (obtained when $\gamma=0$), in addition to gauge invariance. This theory has remarkable integrability properties as can be seen from the fact that coupling it to gravity, led to the first accelerated black holes supported by non-linear electrodynamics \cite{uyui} (see \cite{Flores-Alfonso:2020nnd}-\cite{Barrientos:2024umq} for other interesting self-gravitating configurations).  In flat spacetime, the magnetic Gauss law and Faraday law 
    \begin{align}\label{ecs1}
    	\nabla\cdot\bm{B}=0 \ ,&& \nabla\times\bm{E}+\frac{\partial\bm{B}}{\partial t}=0\ ,
    \end{align}
    are supplemented by the non-linear extensions of the electric Gauss law and the Ampere-Maxwell law, which, in the covariant formulation, are as follows
    \begin{align}\label{ecs2}
    	0&=\partial_{\mu}\left[\cosh\gamma\:F^{\mu\nu}+\sinh\gamma\frac{S\:F^{\mu\nu}+P\:\bar{F}^{\mu\nu}}{\sqrt{S^{2}+P^{2}}}\right]\ .
    \end{align}
    The conformal invariance of the theory in flat spacetime can be seen as a direct consequence of the invariance under Weyl conformal rescaling of the density $\sqrt{-g}\mathcal{L}$. When the theory is formulated on a fixed background, the density $\sqrt{-g}\mathcal{L}$ is invariant under $g_{\mu\nu}(x)\rightarrow\Omega^2(x)g_{\mu\nu}(x)$ and $A_\mu(x)\rightarrow A_\mu(x)$, where $\Omega(x)$ is an arbitrary non-vanishing function of the fixed spacetime with the metric $g_{\mu\nu}(x)$. The symmetry of the field equations under electric-magnetic duality follows as in the original reference \cite{uyui}, which can be seen in a transparent manner in its Pleba\'nski formulation as recently shown in \cite{emod}, as well as in the manifestly duality-symmetric formulation \cite{so2}, that can be also extended to $p-$form in higher dimensions \cite{pforms}\footnote{ModMax theory and it ModMax-Born-Infeld can also be formulated as a four-dimensional extension of the two-dimensional root-$T\bar{T}$ deformation, fulfilling a stress tensor flow equation as shown in \cite{liam1,liam2}}. 
    Let us recall here the basic fact that the Faraday tensor and its dual, in Cartesian coordinates $x^{\mu}=\left(x^0,\vec{x}\right)$ with $x^0=ct$ and mostly plus signature, are explicitly given by
    \begin{align}
    F_{0i}&=-E_i/c\ ,\ F_{ij}=\epsilon_{ijk}B^{k}\ ,\\
    \bar{F}_{0i}=(\star F)_{0i}&=B_i \ ,\ \bar{F}_{ij}=(\star F)_{ij}=\epsilon_{ijk}E_k/c\ .
    \end{align}
    As in the case of Maxwell's theory \cite{d2},  and indeed of any Lorentz invariant theory \cite{Henneaux:2021yzg}, ModMax electrodynamics has two possible Carrollian limits, namely the electric and the magnetic one. We follow the approach of the original reference \cite{ModMax} and work at the level of the field equations.
    
    \subsection{Electric Carrollian limit of ModMax}
    
    In order to take the electric limit it is useful to introduce the rescaling $\bm{E_{e}}=\bm{E}/c$, $s=\left(cC\right)t$ and $\bm{B_{e}}=\left(cC\right)\bm{B}$ in \eqref{ecs1} and \eqref{ecs2}, and then take the limit $C\rightarrow\infty$. Such procedure leads to
    \begin{align}
    	\nabla\times\bm{E_{e}}+\frac{\partial\bm{B_{e}}}{\partial s}=0 \ , && \nabla\cdot\bm{B_{e}}=0 \ ,\\
    	\left(\cosh\gamma+\sinh\gamma\right)\frac{\partial\bm{E_{e}}}{\partial s}=0 \ , && \left(\cosh\gamma+\sinh\gamma\right)\nabla\cdot\bm{E_{e}}=0 \ .
    \end{align}
    As a consequence, the electric Carrolian limit of ModMax, coincides with the mentioned limit of the linear, Maxwell theory. This was already anticipated in \cite{borncar} (see footnote in page two of that reference). Just for completeness, let us mention that these equations are invariant under spacetime translations, space rotations and Carrollian boosts \cite{d2}. The action of the latter on the electromagnetic fields reads, 
    \begin{align}
    	\bm{E_{e}}(\bm{x},s)&\rightarrow\bm{E'_{e}}(\bm{x},s)=\bm{E_{e}}(\bm{x},s-\bm{b}\cdot\bm{x}) \  , \\
    	\bm{B_{e}}(\bm{x},s)&\rightarrow\bm{B'_{e}}(\bm{x},s)=\bm{B_{e}}(\bm{x},s-\bm{b}\cdot\bm{x})-\bm{b}\times\bm{E_{e}}(\bm{x},s-\bm{b}\cdot\bm{x}) \ .
    \end{align}
    
    \subsection{Magnetic Carrollian limit of ModMax}
    
    Following \cite{d2}, the magnetic Carrollian limit can be obtained from the re-scaling $\bm{E_{m}}=C\bm{E}/c$, $\bm{B_{m}}=c\bm{B}$ and $s=\left(cC\right)t$ in \eqref{ecs1} and \eqref{ecs2}. Taking first the $C\rightarrow\infty$ limit on the ModMax Ampere-Maxwell law one obtains
    \begin{equation}
    e^{-\gamma}\left(\nabla\times\bm{B_{m}}-\frac{\partial\bm{E_{m}}}{\partial s}\right)-2\sinh\gamma\frac{\bm{B_{m}}\cdot\dfrac{\partial\bm{E_{m}}}{\partial s}}{B_{m}^{2}}\bm{B_{m}}=0
    \end{equation}
    These equations can be further simplified by taking its scalar product with the magnetic field $\bm{B_{m}}$, which implies the identity $-e^{\gamma}\bm{B_{m}}\cdot\frac{\partial\bm{E_{m}}}{\partial s}=0$. Consequently, the magnetic Carrollian limit of ModMax theory leads to the system
    \begin{align}\label{ecam1}
    	\frac{\partial\bm{B_{m}}}{\partial s}=0\ ,&& e^{-\gamma}\left(\nabla\times\bm{B_{m}}-\frac{\partial\bm{E_{m}}}{\partial s}\right)=0 \, \\ \label{ecam2}
     \nabla\cdot\bm{B_{m}}=0\ , && e^{-\gamma}\:\nabla\cdot\bm{E_{m}}+2\sinh\gamma\:\left(\bm{B_{m}}\cdot\nabla\right)\frac{\bm{B_{m}}\cdot\bm{E_{m}}}{B_{m}^{2}}=0 \ .
    \end{align}
    Interestingly, the only ModMax equation that is sensitive to the nonlinearity of the theory in the magnetic Carrollian limit is the ModMax Gauss law. When $\gamma=0$, the system reduces to the magnetic Carrollian Maxwell system originally found in \cite{d2}. It is evident that the system of equations is invariant under spacetime translations and spatial rotations, and it can be checked that the system is invariant under the magnetic Carrollian boosts
    \begin{align}
    	\bm{B_{m}}(\bm{x},s)&\rightarrow\bm{B'_{m}}(\bm{x},s)=\bm{B_{m}}(\bm{x},s-\bm{b}\cdot\bm{x})\\
    	\bm{E_{m}}(\bm{x},s)&\rightarrow\bm{E'_{m}}(\bm{x},s)=\bm{E_{m}}(\bm{x},s-\bm{b}\cdot\bm{x})+\bm{b}\times\bm{B_{m}}(\bm{x},s-\bm{b}\cdot\bm{x})\ .
    \end{align}
    From this analysis we conclude that the system \eqref{ecam1} and \eqref{ecam2} does indeed define a Carrollian field theory, which is naturally derived from the magnetic $c\rightarrow 0$ limit of ModMax theory. In other words, we have shown how ModMax electrodynamics allows for a non-trivial Carrollian limit. Before continuing, it is interesting to notice that the invertible transformation $(\bm{E_m},\bm{B_m})\rightarrow(\mathfrak{E},\mathfrak{B})$ and their explicit inverse defined by 
    \begin{align}\label{tra1}
    	\mathfrak{E}&=\bm{E_{m}}+2e^{\gamma}\sinh\gamma\frac{\bm{B_{m}}\cdot\bm{E_{m}}}{B_{m}^{2}}\bm{B_{m}} \ , & \mathfrak{B}&=\bm{B_{m}} \ ,  \\ \label{tra2} 
    	\bm{E_{m}}&=\mathfrak{E}-2e^{-\gamma}\sinh\gamma\:\frac{\mathfrak{E}\cdot\mathfrak{B}}{\mathfrak{B}^{2}}\mathfrak{B}\ , & \bm{B_{m}}&=\mathfrak{B} \ ,
    \end{align}
    maps the system \eqref{ecam1} and  \eqref{ecam2} to Maxwell's magnetic Carrollian equations, namely $\mathfrak{E}$ and $\mathfrak{B}$ fulfill 
    \begin{align}
    	\frac{\partial\mathfrak{B}}{\partial s}&=0 \ ,& \nabla\cdot\mathfrak{B}&=0 \ ,\\
    	\nabla\times\mathfrak{B}-\frac{\partial\mathfrak{E}}{\partial s}&=0  \ , & \nabla\cdot\mathfrak{E}&=0\ .
    \end{align}
    The transformations \eqref{tra1} and \eqref{tra2} must be interpreted as a duality between magnetic Carrollian ModMax and magnetic Carrollian Maxwell theories, which allows to map vacuum solutions from one theory to the other. This duality by no means trivializes the magnetic Carrollian limit of the ModMax theory, since even though the solution spaces are connected in vacuum, the duality might be broken as soon as the field theories are coupled to a generic Carrollian charged source. Namely, the backreaction of a Carrollian source will be different in the two theories. The map is also ill-defined when considering fields with singularities. 
    
\begin{figure}[h!]
\begin{center}
\includegraphics[scale=0.45]{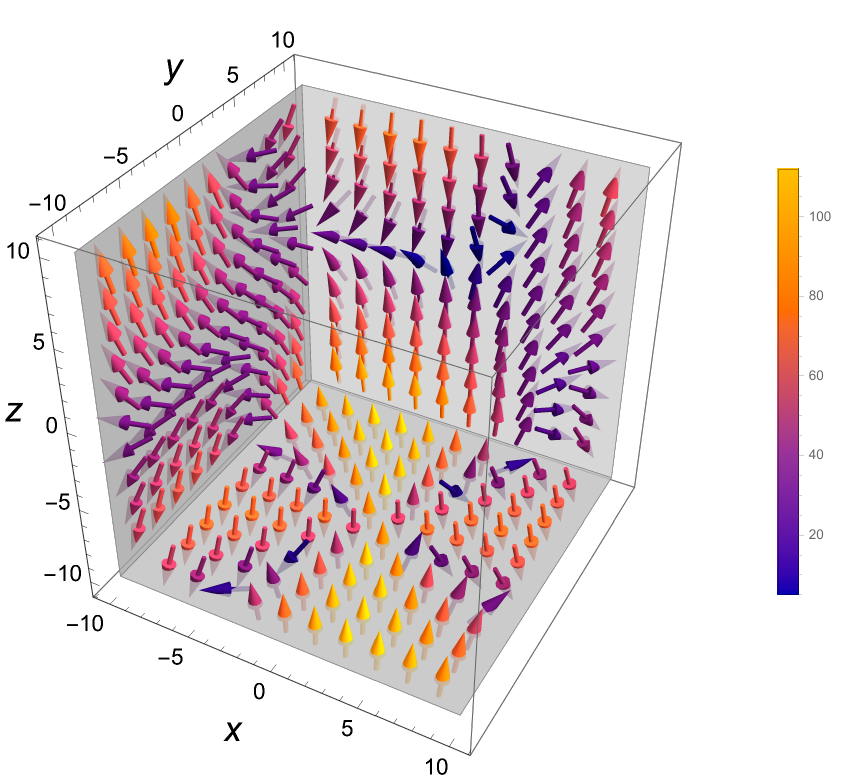}
\includegraphics[scale=0.45]{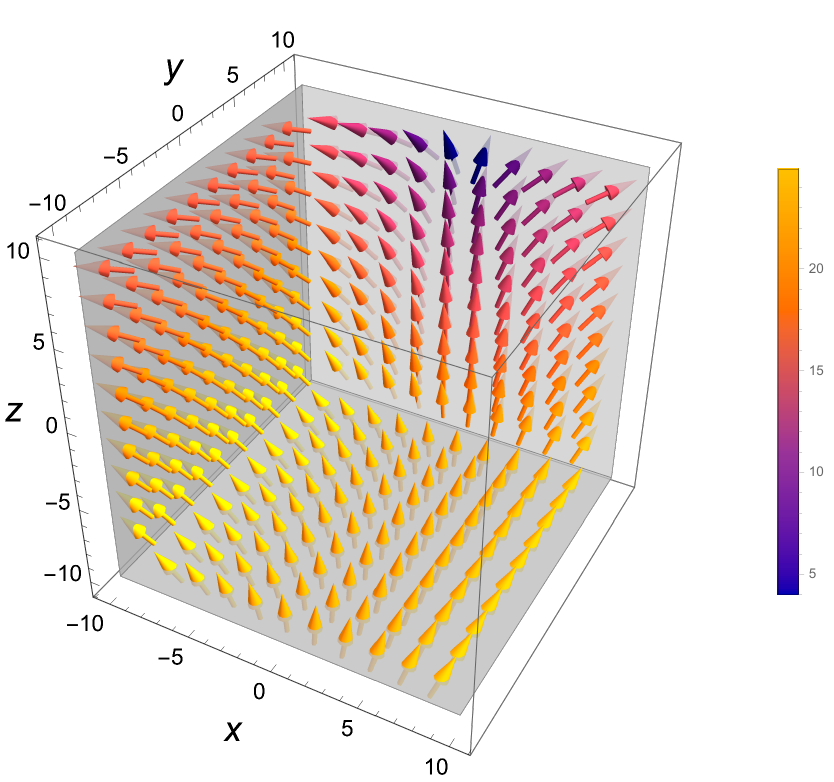}
\includegraphics[scale=0.45]{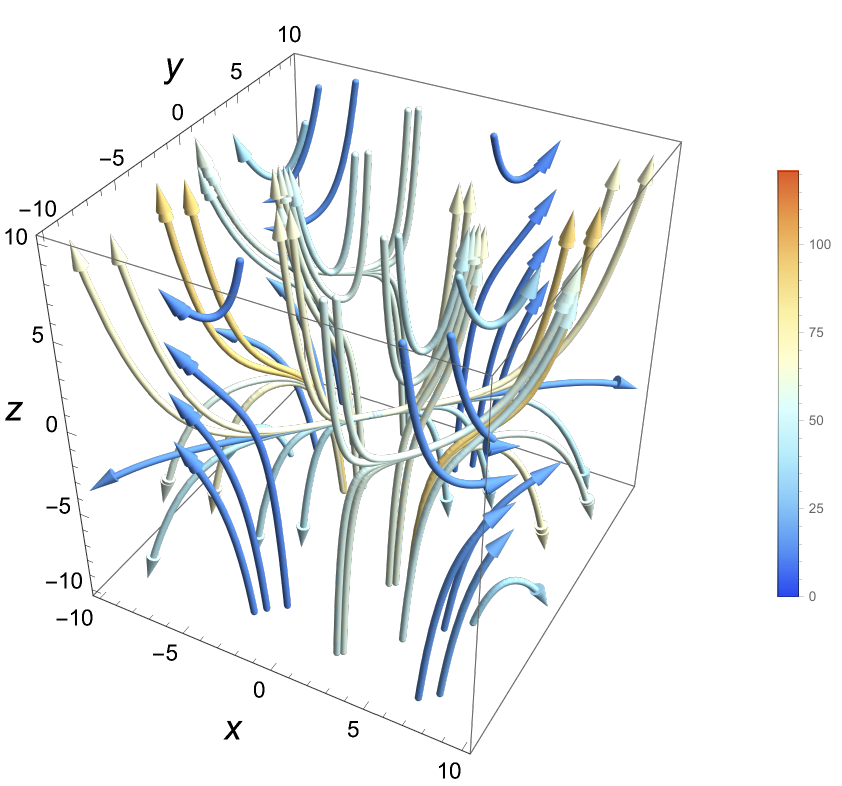}
\includegraphics[scale=0.45]{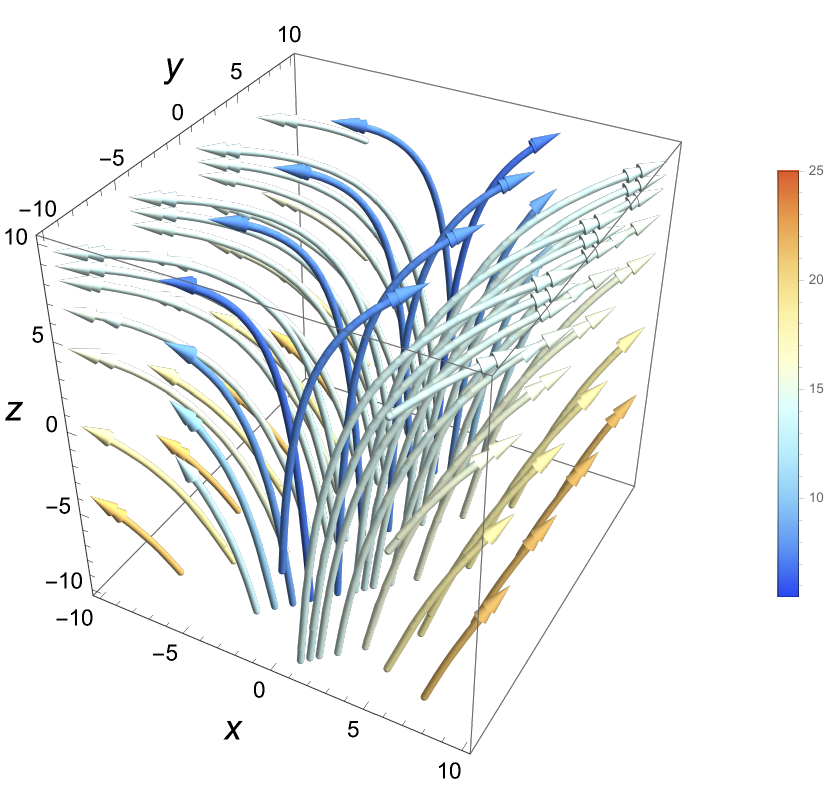}  
\end{center}
\caption{Upper panel: The left panel shows the vector field of the solution \eqref{electricsolution} for $\gamma=1$, while the right panel shows the case $\gamma=0$. Bottom panel: The stream plot is shown in the left (right) panel for the ModMax $\gamma=1$ (Maxwell $\gamma=0$) case. In all figures we set $s=6$ and $E_0=1=B_0$.}
\label{figures}
\end{figure}

    In order to explore the consequences of the magnetic Carrollian ModMax theory defined by the equations \eqref{ecam1} and \eqref{ecam2}, it is useful to study the space of solutions. We consider the simplest non-trivial, divergenceless magnetic field $\bm{B_m}=B_0(-y,x,0)$, referred to its Cartesian components, where $B_0$ is a constant. Manipulating the remaining equations, one can check that the system is solved when the electric field is given by
    \begin{equation}
    \bm{E_m}=\left(xE_0,xE_0,2sB_0-\frac{2xz(x+y)-e^{2\gamma}z(x^2+2xy-y^2)}{x^2+y^2}E_0+F(x,y)\right)\ ,\label{electricsolution}
    \end{equation}
where $E_0$ is a constant, $F(x,y)$ is an arbitrary function and note the non-trivial contribution from the ModMax coupling $\gamma$. The presence of arbitrary functions in the gauge invariant fields is not surprising in the present context, since the field equations possess invariance under supertranslations generated by arbitrary space-dependent functions (see next section). In the Maxwell limit, namely when $\gamma\rightarrow 0$ and $F(x,y)=0$, the electric field takes a particularly simple form $\bm{E_m^{\gamma=0}}=(xE_0,xE_0,2sB_0-zE_0)$. Figure \ref{figures} shows the behavior of the magnetic Carrollian ModMax solution and contrasts it with the $\gamma=0$ case. One can clearly see that the presence of the singularity at $x=y=0$ is removed in the Maxwell limit.
In the next section we study the invariances of the magnetic Carrollian ModMax system, using the transformations \eqref{tra1} and \eqref{tra2} in our favor, from the point of view of Lie point symmetries. Such an analysis leads to an infinite dimensional extension of the global Carrollian group.
    
\section{Lie point symmetries of the Carrollian ModMax theory}\label{three}
Since only the magnetic Carrollian limit of ModMax theory is sensitive to the nonlinear terms in the field equations, we will focus on studying the Lie point symmetries of such a system. As it is known \cite{liep1,liep2}, the Lie point symmetries are transformations which map solutions to solutions of the field equations acting both on the coordinates $(s,\bm{x})$ and on the fields $(\mathbf{E},\mathbf{B})$. One defines an extended manifold $(s,\bm{x},\mathbf{E},\mathbf{B})$ and consistently proposes a generic transformation whose infinitesimal version is generated by a vector living on the tangent space of the extended manifold. Imposing that the infinitesimal transformation maps solutions into solutions, restricts that form of the generators leading to the symmetry transformations of the system. Proceeding in this way,  the symmetries for the magnetic Carrollian ModMax theory \eqref{ecam1}-\eqref{ecam2} are written in the Table \ref{table:1} \footnote{We have used the Mathematica Package from the book \cite{liep2}.}.
    \renewcommand{\arraystretch}{0.8}
    \begin{table}[h!]
    \begin{center}
    \begin{tabular}{c|c|}
    \cline{2-2}
                           & $(s,\bm{x},\bm{E},\bm{B})\rightarrow$ \\ \hline
    \multicolumn{1}{|c|}{Space-time translations} & $(s+\lambda,\bm{x}+\bm{\lambda},\bm{E},\bm{B})$ \\ [1ex] \hline
    \multicolumn{1}{|c|}{Rotations} & $(s,R\bm{x},R\bm{E},R\bm{B})$ \\ [1ex] \hline
    \multicolumn{1}{|c|}{Supertranslations} & $(s+f(\bm{x}),\bm{x},\bm{E}-\nabla f\times\bm{B},\bm{B})$ \\ [1ex] \hline
    \multicolumn{1}{|c|}{Space dilations} & $(s,e^{\lambda}\bm{x},\bm{E},e^{\lambda}\bm{B})$ \\ [1ex] \hline
    \multicolumn{1}{|c|}{SCT along $x$ } & $ \left(\omega_{x}(\lambda)s,\omega_{x}(\lambda)\left(x{-}\lambda\bm{x}^2\right),\omega_{x}(\lambda)y,\omega_{x}(\lambda)z,T_{1}(\lambda)\bm{E}{+}O_{1}(\lambda)\bm{B},T_{1}(\lambda)\bm{B}\right) $ \\ [1ex] \hline
    \multicolumn{1}{|c|}{SCT along $y$ } & $ \left(\omega_{y}(\lambda)s,\omega_{y}(\lambda)x,\omega_{y}(\lambda)\left(y{-}\lambda\bm{x}^2\right),\omega_{y}(\lambda)z,T_{2}(\lambda)\bm{E}{+}O_{2}(\lambda)\bm{B},T_{2}(\lambda)\bm{B}\right) $ \\ [1ex] \hline
    \multicolumn{1}{|c|}{SCT along $z$ } & $ \left(\omega_{z}(\lambda)s,\omega_{z}(\lambda)x,\omega_{z}(\lambda)y,\omega_{z}(\lambda)\left(z{-}\lambda\bm{x}^2\right),T_{3}(\lambda)\bm{E}{+}O_{3}(\lambda)\bm{B},T_{3}(\lambda)\bm{B}\right) $\\ [1ex] \hline
    \multicolumn{1}{|c|}{Field dilations} & $ (s,\bm{x},e^{\lambda}\bm{E},e^{\lambda}\bm{B})$ \\ [1ex] \hline
    \multicolumn{1}{|c|}{Generalized EM duality} & $ (s,\bm{x},\bm{E}-\lambda e^{-2\gamma}\bm{B},\bm{B})$ \\ [1ex] \hline
    \end{tabular}
    \end{center}
    \caption{Action of the symmetries on the fields. Here SCT stands for special conformal translations, $\lambda$ is a real number, $\bm{\lambda}$ is a vector, $R\in SO(3)$, $f$ is a differentiable function of the spatial coordinates $\bm{x}$ only, and $\omega_i(\lambda)$, $T_i(\lambda)$ and $O_i(\lambda)$ are defined in the appendix.}\label{table:1}
    \end{table}
    
    The infinitesimal action of the spacetime transformations are implemented by the differential operators
    \begin{align}
    	P_{A}&=\frac{\partial}{\partial x^{A}}\, , & H=&\frac{\partial}{\partial s} \, ,  & J_{A}&=\epsilon_{ABC}x^{B}\frac{\partial}{\partial x^{C}}\, ,\\ 
     T_{\ell m n}&=x^{\ell}y^{m}z^{n}\frac{\partial}{\partial s} \, , 
     & D=&x^{A}\frac{\partial}{\partial x^{A}}\, ,&  Q&=s\frac{\partial}{\partial s}\, , \label{losopdif} 
    \end{align}
    \begin{equation}
    S_{A}=2x_{A}\left(x^{B}\frac{\partial}{\partial x^{B}}+s\frac{\partial}{\partial s}\right)-x_{B}x^{B}\frac{\partial}{\partial x^{A}}
    \end{equation}
    with $A\in\{1,2,3\}$ and $n,m,k\in\mathbb{N}_{0}$. As expected, defining
    \begin{align}
    	K_{1}&=T_{100}=x\frac{\partial}{\partial s} & K_{2}&=T_{010}=y\frac{\partial}{\partial s} & K_{3}&=T_{001}=z\frac{\partial}{\partial s}
    \end{align}
    one obtains that the generators $\left\{P_{A},H,J_{A},K_{A}\right\}$ realize the Carroll Algebra
    \begin{align}
    	\left[P_{A},P_{B}\right]&=0 & \left[P_{A},H\right]&=0 & \left[J_{A},P_{B}\right]&=-\epsilon_{ABC}P_{C}\nonumber\\
    	\left[P_{A},K_{B}\right]&=\delta_{AB}H &  \left[K_{A},K_{B}\right]&=0 & \left[H,K_{A}\right]&=0\\
    	\left[J_{A},J_{B}\right]&=-\epsilon_{ABC}J_{C} & \left[J_{A},K_{B}\right]&=-\epsilon_{ABC}K_{C} & \left[H,J_{A}\right]&=0\nonumber
    \end{align}
    
    The enlarged set of symmetry generators $\left\{P_{A},H,J_{A},D,Q,S_A,T_{\ell m n}\right\}$, containing the whole family of spacetime symmetries, including the general supertranslations, satisfy the following commutators
    \begin{align}\notag
    	\left[D, P_{A}\right]&=-P_{A} \, ,  & \left[D,H\right]&=0 \, ,& \left[D,J_{A}\right]&=0  \, , \\ \notag
    		\left[D, Q\right]&=0\, ,  & \left[D,S_{A}\right]&=S_{A}\, ,& \left[D,T_{\ell m n}\right]&=\left(\ell +m +n\right)T_{\ell m n}  \, , \\ \notag
    		\left[H, Q\right]&=H \, ,  & \left[H,S_{A}\right]&=2K_{A} \, ,&  \left[H,T_{\ell m n}\right]&=0 \, , \\
    		\left[Q, P_{A}\right]&=0\, ,  &  \left[Q, J_{A}\right]&=0\, ,  & \left[Q, S_{A}\right]&=0\, , \\ \notag
    		 \left[S_{A},S_{B}\right]&=0\, , & \left[J_{A},S_{B}\right]&=-\epsilon_{ABC}S_{C}
    & \left[P_{A},S_{B}\right]&=2\delta_{AB}\left(D+Q\right)-2 \epsilon_{ABC}J_C \, , \\ \notag
    	\left[T_{\ell m n},T_{rst}\right]&=0 \, , & \left[T_{\ell m n},Q\right]&=T_{\ell m n} \, , \\ \notag
    	\left[T_{\ell m n},P_{1}\right]&=-\ell \, T_{\ell-1\, m\, n}\, , & \left[T_{\ell m n},P_{2}\right]&=-m\, T_{\ell\, m-1\, n} \, , & \left[T_{nmj},P_{3}\right]&=-n\,  T_{\ell\, m\, n-1}\, ,\notag
    \end{align}
\vspace*{-1.5cm}
    \begin{align}
    	\left[T_{\ell m n},S_{1}\right]&=\left(2-\ell-2m-2n\right)T_{\ell+1\, m\,n}+\ell \left(T_{\ell-1\, m+2\, n}+T_{\ell-1\, m\, n+2}\right)\nonumber\\
    	\left[T_{\ell m n},S_{2}\right]&=\left(2-2\ell-m-2n\right)T_{\ell \,m+1\,n}+m\left(T_{\ell+2\, m-1\, n}+T_{\ell\, m-1\,n+2}\right)\nonumber\\
    	\left[T_{\ell m n},S_{3}\right]&=\left(2-2\ell-2m-n\right)T_{\ell\, m\, n+1}+n\left(T_{\ell\,m+2\,n-1}+T_{\ell+2\,m\,n-1}\right)\\
\left[J_{1},T_{\ell m n}\right]&=n\, T_{\ell\,m+1\,n-1}-m\, T_{\ell\,m-1\,n+1}\nonumber\\
    	\left[J_{2},T_{\ell m n}\right]&=\ell \,T_{\ell-1\, m\, n+1}-n\, T_{\ell+1\,m\,n-1}\nonumber\\
    	\left[J_{3},T_{\ell m n}\right]&=m\, T_{\ell+1\,m-1\,n}-\ell \, T_{\ell-1\,m+1\,n}\nonumber
    \end{align}
    Consistently with the defined range of $\{\ell, m, n\}$ is $\mathbb{N}_{0}^{3}$, the downward ladder is truncated at zero for the index values of $T_{\ell m n}$. To characterize this algebra, it is useful to review some basic facts about weak flat Carrollian structures, which are defined by a manifold $M=\mathbb{R}_s\times\mathbb{R}^d$ with Cartesian coordinates $(s,x^A)$, a degenerate metric $g=g_{AB}dx^A\otimes dx^B$, and a vector field $\xi=\frac{\partial}{\partial s}$. The conformal Carroll groups of level $k$ are defined by the vector fields $X$ such that \cite{d2}
    \begin{align}
    	\label{eq:conf-carr-crit}
    	L_{X}\left(g\otimes\xi^{\otimes k}\right)=0
    \end{align}
    where $L_X$ stands for the Lie derivative along the vector field $X$. In the special case $k=2$, which turns out to be relevant for this work, one has explicitly that $X$ belongs to the Conformal Carroll algebra of level $2$, provided that
    \begin{align}
    	&0=L_{X}\left(g\otimes\xi^{\otimes 2}\right)=\label{confrest}\\ \notag
     &\delta_{AB}\left(L_{X}dx^{A}\right)\otimes dx^{B}\otimes\xi^{\otimes 2}+\delta_{AB}dx^{A}\otimes\left(L_{X}dx^{B}\right)\otimes\xi^{\otimes 2}+g\otimes\left(L_{X}\xi\right)\otimes\xi+g\otimes\xi\otimes\left(L_{X}\xi\right)\ .
    \end{align}
    Consequently, to evaluate whether or not $X$ belongs to the Conformal Carroll algebra of level $2$, we must compute
    \begin{align}
L_{X}dx^{A}:=i_Xd\left(dx^A\right)+di_X\left(dx^A\right)=d\left(dx^{A}\left(X\right)\right) \ \ \ \text{and}\ \ \   L_{X}\xi:=\left[X,\xi\right]\ .
    \end{align}
    It can be shown directly that $L_{P_I}dx^A=L_Hdx^A=0$ as well as $L_{P_I}\xi=L_H\xi=L_{J_I}\xi=0$, which together with the fact that $L_{J_I}dx^A=\epsilon_{IJA}dx^J$ is antisymmetric, leads to $L_X\left(g\otimes\xi\otimes\xi\right)=0$, and thus, as expected, belongs to the Conformal Carrollian algebra of level $2$. The super-translation generators are in the kernel of $dx^{A}$ and also have a vanishing commutator with $\xi$, namely $L_{T_{mnk}}dx^A=0=L_{T_{mnk}}\xi$, so as already known, the generators $T_{mnk}$ also belong to the conformal Carroll algebra of level $2$. The generators $X=S_A$ fulfill \eqref{confrest} in a non-trivial manner, since $ 
L_{S_{A}}dx^{I}=2x^{I}dx_{A}+2x_{A}dx^{I}-2\delta^{I}_{A}x_{B}dx^{B}$, and $L_{S_{A}}\xi=-2x_{A}\xi$, from where it follows that $L_{S_{A}}\left(g\otimes\xi^{\otimes 2}\right)=0$.
It can be checked also that $L_Ddx^A=dx^A$ and $L_D\xi=0$, while $L_Qdx^A=0$ and $L_Q\xi=-\xi$, namely $D$ and $Q$ independently, do not belong to the Conformal Carrollian algebra of level 2. Interestingly, however, the diagonal symmetry subgroup of the magnetic Carrollian ModMax theory generated by $Q+D$ do belong to the conformal Carrollian algebra of level $2$, namely $L_{Q+D}\left(g\otimes\xi^2\right)=0$. The consistency of our analysis can be confirmed by recalling that equation (III.15) of reference \cite{conformal} gives the explicit expression of the general vector field generating the conformal Carrollian algebra of level $k$, namely
    \begin{align}
    X=&\left(\omega^A_Bx^B+\gamma^A+\chi x^A+\kappa^Ax_Bx^B-2\kappa_Bx^Bx^A\right)\frac{\partial}{\partial x^A}\nonumber\\
    &\left(\frac{2}{k}\left(\chi-2\kappa x^A\right)s+F\left(x^A\right)\right)\frac{\partial}{\partial s}\in\mathfrak{ccarr}_k(d+1)
    \end{align}
    where the parameter $\chi$ is the same for the spatial and time dilations when $k=2$, consistently with the fact that neither $Q$ nor $D$ belong to $\mathfrak{ccarr}_k(d+1)$, but $Q+D$ does.

\section{Conclusions}\label{four} 
We have shown that the magnetic Carrollian limit of ModMax electrodynamics, lead to a non-trivial, nonlinear field theory. The symmetry generators contain a countable infinite family of transformations which act on the Carrollian time as $s\rightarrow s+f(\bm{x})$, where $f$ is an arbitrary function of the spatial coordinates. We also found an exact solution to the system, that makes explicit the non-trivial role of the ModMax $\gamma$ parameter, even after the magnetic Carrollian limit has been taken. A diagonal subgroup of the Lie Point Symmetry group we have identified, span the Conformal Carrollian algebra of level $2$, $\mathfrak{ccarr}_2(3+1)$.

Before finishing, let us notice that the collapse of the light cone in the Carrollian limit of the Minkowski metric has a similar effect on test particles, as the $l\rightarrow0$ limit of the global AdS spacetime metric with curvature radius $l$, in the latter case due to the infinite gravitational potential that a particle must overcome to depart from the origin, which actually would occur at every point due to the homogeneous nature of AdS spacetime. In both cases, a particle at the origin will not be able to move to another point in spacetime. This relation has a nice counterpart in field theory:  recently in \cite{swiftons} it was shown that for different field theories, including scalars, that propagation outside the light cone, which from the point of view of flat spacetime would be associated with tachyonic behavior, can produce an energy bounded from below if it occurs in Carrollian spacetime. The AdS counterpart of this result is given by the fact that on such a spacetime, fields with negative squared mass can lead to positive energy, provided that $m^{2}$ is bounded from below by a negative number which depends on both the dimension and the spin of the relativistic field theory. Such a bound, known as the Breitenlohner-Freedman bound, generically has the form $m^{2}\geq m_{BF}^{2}=-\frac{n\left( D,s\right) }{l^{2}}$, which goes to minus infinity in the AdS ultrastatic limit $l\rightarrow 0$, mimicking the structure of the Swiftons of Carrollian physics. It would be interesting to further explore this relation.

\vspace{-0.3cm}

    \section*{Acknowledgments}
    We thank Patrick Concha, Mokhtar Hassaine and Evelyn Rodriguez, for enlightening comments. This work is partially funded by FONDECYT Regular grants 1211356 and 1221504. A.H. also thanks the support of Dirección de Postgrado UdeC. FC would  like to thank Universidad Austral de Chile and Universidad de Concepci\'on where this work originally started.

\section*{Appendix}
In this appendix we provide the definitions of $\omega_i(\lambda)$, $O_i(\lambda)$ and $T_i(\lambda)$, introduced in Table I:
\begin{align}
	\omega_{x}(\lambda)&=\frac{x^{2}+y^{2}+z^{2}}{\left(x-\lambda\left(x^{2}+y^{2}+z^{2}\right)\right)^{2}+y^{2}+z^{2}}\\
	\omega_{y}(\lambda)&=\frac{x^{2}+y^{2}+z^{2}}{x^{2}+\left(y-\lambda\left(x^{2}+y^{2}+z^{2}\right)\right)^{2}+z^{2}}\\
	\omega_{z}(\lambda)&=\frac{x^{2}+y^{2}+z^{2}}{x^{2}+y^{2}+\left(z-\lambda\left(x^{2}+y^{2}+z^{2}\right)\right)^{2}}
\end{align}
For convenience, let us also define the accompanying factors
\begin{align}
	\Omega_{x}(\lambda)&=\left(\lambda x-1\right)^{2}+\lambda^{2}\left(y^{2}+z^{2}\right)\\
	\Omega_{y}(\lambda)&=\left(\lambda y-1\right)^{2}+\lambda^{2}\left(x^{2}+z^{2}\right)\\
	\Omega_{z}(\lambda)&=\left(\lambda z-1\right)^{2}+\lambda^{2}\left(x^{2}+y^{2}\right)\ ,
\end{align}
Using these factors we can write two families of matrices that characterize the action of special conformal Carrollian transformations on the electric and magnetic field. The first family of matrices is $T_{i}(\lambda)$, is given by
{\footnotesize{
\begin{align}
	T_{x}(\lambda)&=\Omega_{x}(\lambda)\begin{pmatrix}
		\lambda  \left(\lambda  x^2-2 x-\lambda  \left(y^2+z^2\right)\right)+1 & 2 \lambda  y (\lambda  x-1) & 2 \lambda  z (\lambda  x-1) \\
		-2 \lambda  y (\lambda  x-1) & \lambda ^2 \left(x^2-y^2+z^2\right)-2 \lambda  x+1 & -2 \lambda ^2 y z \\
		-2 \lambda  z (\lambda  x-1) & -2 \lambda ^2 y z & \lambda ^2 \left(x^2+y^2-z^2\right)-2 \lambda  x+1 \\
	\end{pmatrix}\\
	T_{y}(\lambda)&=\Omega_{y}(\lambda)\begin{pmatrix}
		\lambda ^2 \left(-x^2+y^2+z^2\right)-2 \lambda  y+1 & -2 \lambda  x (\lambda  y-1) & -2 \lambda ^2 x z \\
		2 \lambda  x (\lambda  y-1) & \lambda  \left(-\lambda  \left(x^2+z^2\right)+\lambda  y^2-2 y\right)+1 & 2 \lambda  z (\lambda  y-1) \\
		-2 \lambda ^2 x z & -2 \lambda  z (\lambda  y-1) & \lambda ^2 \left(x^2+y^2-z^2\right)-2 \lambda  y+1 \\
	\end{pmatrix}\\
	T_{z}(\lambda)&=\Omega_{z}(\lambda)\begin{pmatrix}
		\lambda ^2 \left(-x^2+y^2+z^2\right)-2 \lambda  z+1 & -2 \lambda ^2 x y & -2 \lambda  x (\lambda  z-1) \\
		-2 \lambda ^2 x y & \lambda ^2 \left(x^2-y^2+z^2\right)-2 \lambda  z+1 & -2 \lambda  y (\lambda  z-1) \\
		2 \lambda  x (\lambda  z-1) & 2 \lambda  y (\lambda  z-1) & \lambda  \left(z (\lambda  z-2)-\lambda  \left(x^2+y^2\right)\right)+1 \\
	\end{pmatrix}
\end{align}
}}
The second family of matrices is $O_{i}(\lambda)$, with each one given by

\begin{align}
	O_{x}(\lambda)&=\Omega_{x}(\lambda)\begin{pmatrix}
		0 & 2 \lambda ^2 s z & -2 \lambda ^2 s y \\
		2 \lambda ^2 s z & 0 & -2 \lambda  s (\lambda  x-1) \\
		-2 \lambda ^2 s y & 2 \lambda  s (\lambda  x-1) & 0 \\
	\end{pmatrix}\\
	O_{y}(\lambda)&=\Omega_{y}(\lambda)\begin{pmatrix}
		0 & -2 \lambda ^2 s z & 2 \lambda  s (\lambda  y-1) \\
		-2 \lambda ^2 s z & 0 & 2 \lambda ^2 s x \\
		-2 \lambda  s (\lambda  y-1) & 2 \lambda ^2 s x & 0 \\
	\end{pmatrix}\\
	O_{z}(\lambda)&=\Omega_{z}(\lambda)\begin{pmatrix}
		0 & -2 \lambda  s (\lambda  z-1) & 2 \lambda ^2 s y \\
		2 \lambda  s (\lambda  z-1) & 0 & -2 \lambda ^2 s x \\
		2 \lambda ^2 s y & -2 \lambda ^2 s x & 0 \\
	\end{pmatrix}
\end{align}

    \end{document}